\documentstyle[12pt,a4]{article}

\begin{document}

\title{Classical states and decoherence by unitary evolution in the thermodynamic limit}
\author{Marco Frasca \\
        Via Erasmo Gattamelata, 3 \\
        00176 Roma (Italy) \\
        e-mail:marcofrasca@mclink.it}

\date{ \ }

\maketitle

\abstract{
It is shown how classical states, meant as states representing a classical object, can be produced
in the thermodynamic limit, retaining the unitary evolution of quantum mechanics. Besides,
using a simple model of a single spin interacting with a spin-bath, it is seen how decoherence, with
the off-diagonal terms in the density matrix going to zero, can be obtained when the
number of the spins in the bath is taken to go formally to infinity. In this case, indeed, the
system appears to flop at a frequency being formally infinity that, from a physical standpoint,
can be proved equivalent to a time average.
}

KEYWORDS: Classical states; Thermodynamic limit; Unitary evolution; Decoherence.

\newpage

The question of how a classical world could emerge from quantum mechanics is
a longstanding problem still deeply debated today. A sound solution has been
offered by decoherence (see e.g. \cite{zur}) that gives also a possible direction
to the solution of the problem of measurement.

Decoherence is based on the fact that whatever quantum system we are going to
consider, this is unavoidably in contact with an environment and this can be
enough to make the system lose the coherence of its quantum evolution. By
construction, it appears that decoherence, defined in this way, is a dissipative
effect.

Quite recently, non-dissipative approaches have also been proposed by Bonifacio
et al. \cite{bon} and Sun et al. \cite{sun}. Indeed, dissipative decoherence
has to cope with the question of a kind of regression to solve the measurement
problem as also the measurement apparatus should decohere to be considered a
classical system, that is, one has at least three levels to have a quantum
system that decoheres and give a pointer on an instrument to indicate a definite
value.

Our aim here is to discuss a recent proposal to obtain classical states from quantum unitary
evolution \cite{fra}. This is based on the formal analogy between quantum
mechanics and statistical mechanics (a Wick rotation) that can make the
classical world emerge in the ``thermodynamic limit'' as happens to thermodynamics.

This approach, when applied to a simple model devised by Zurek \cite{zur2} of
a spin in a spin-bath, proves to give rise to ordinary decoherence, that is,
erasing of the off-diagonal terms in the density matrix of the spin. In this
case one has a spin flopping with a frequency proportional to the number of spins
in the bath and this means, in the thermodynamic limit, an infinite frequency.
Physically, this is proved to be equivalent to an average in time. Mathematically,
this is a typical example of a divergent series \cite{hardy}. Anyhow, it
is essential to emphasize that unitary evolution is retained at each step of
our analysis.

An example of a classical system obtained by unitary evolution 
in the thermodynamic limit is given by ($\hbar=1$)
\begin{equation}
    H=\lambda\sum_{i=1}^N\sigma_{zi}
\end{equation}
that is, a set of $N$ two-level systems. It can be proved that, when
the system is prepared with the initial wave-function
\begin{equation}
    |\psi(0)\rangle=\prod_{i=1}^N(\alpha_i|\downarrow\rangle_i+\beta_i|\uparrow\rangle_i) \label{eq:state}
\end{equation}
being $\sigma_{zi}|\uparrow\rangle_i=|\uparrow\rangle_i$ and 
$\sigma_{zi}|\downarrow\rangle_i=-|\downarrow\rangle_i$, then the quantum fluctuations
of the Hamiltonian can be neglected as
\begin{equation}
    \frac{\Delta H}{\langle H\rangle}\propto \frac{1}{\sqrt{N}}
\end{equation}
in the thermodynamic limit $N\rightarrow\infty$. Then, the spins $\sum_{i=1}^N\sigma_{xi}$
and $\sum_{i=1}^N\sigma_{yi}$ follows the classical equations of motion without any
deviation due to quantum fluctuations. These are the equations for the mean values
as given by the Ehrenfest theorem.

So, this system, with a properly prepared initial state, behaves classically. Then, it
is also able to induce decoherence \cite{zur2}: The initial state (\ref{eq:state})
appears as a generally realistic state for a set of properly disordered two-level
systems. An important conclusion from the above example is that such a classical
object obtained by unitary evolution depends on the way the system is 
initially prepared.

Having proved that a set of two-level systems can behaves as a classical system
in its quantum evolution, we want to show how it can induce decoherence on a
single mode of radiation field.

The interaction of N two-level systems with a single radiation mode is given by
\begin{equation}
    H=\Delta\sum_{i=1}^N\sigma_{zi}+\omega a^\dagger a+g(a+a^\dagger)\sum_{i=1}^N\sigma_{xi}
\end{equation}
with no rotating wave approximation. Indeed, we aim to study this model in the
strong coupling regime. That is, we want to apply a dual Dyson series \cite{fra1}
\begin{equation}
    U_D(t)=U_F(t)T\exp\left[-i\int_0^tdt'U_F^\dagger(t')H_0 U_F(t)\right]
\end{equation}
being $T$ the time-ordering operator with
\begin{equation}
    H_0=\Delta\sum_{i=1}^N\sigma_{zi}
\end{equation}
and
\begin{equation}
    \left[\omega a^\dagger a+g(a+a^\dagger)\sum_{i=1}^N\sigma_{xi}\right]U_F(t)=
	i\frac{\partial}{\partial t}U_F(t).
\end{equation}
The solution at the leading order, a non trivial one, is then written as
\begin{equation}
    |\psi(t)\rangle\approx U_F(t)|\psi(0)\rangle.
\end{equation}

The unitary evolution to evaluate the perturbation series is easily written
down as \cite{fra}
\begin{equation}
    U_F(t)=e^{i\hat\xi(t)}e^{-i\omega a^\dagger at}\exp[\hat\alpha(t)a^\dagger-\hat\alpha^*(t)a]
\end{equation}
being
\begin{equation}
    \hat\xi(t)=\frac{\left(\sum_{i=1}^N\sigma_{xi}\right)^2g^2}{\omega^2}(\omega t-\sin(\omega t))
\end{equation}
and
\begin{equation}
    \hat\alpha(t)=\frac{\left(\sum_{i=1}^N\sigma_{xi}\right)g}{\omega}(1-e^{i\omega t}).
\end{equation}
Assuming the initial state being $|\psi(0)\rangle=|0\rangle\prod_{i=1}^N|-1\rangle_i$ where
the environment is taken to be in eigenstates of $\sigma_{xi}$ and the field
in the ground state, we obtain the leading order solution
\begin{equation}
    |\psi(t)\rangle\approx
	e^{i\frac{N^2g^2}{\omega^2}(\omega t -\sin(\omega t))}
	|\alpha(t)\rangle \prod_{i=1}^N|-1\rangle_i 
\end{equation}
having set
\begin{equation}
    \alpha(t)=\frac{Ng}{\omega}(1-e^{-i\omega t})
\end{equation}
that means that at the leading order the field evolves as a coherent state. This
in turns means that, in the thermodynamic limit $N\rightarrow\infty$, the radiation
mode is described by a classical field\cite{mand}.

It is interesting to note that the environment is left untouched at this order.

The first order correction is easily written down as
\begin{equation}
    |\psi^{(1)}(t)\rangle=-iU_F(t)\int_0^tdt'U_F^\dagger(t')
	\Delta\sum_{i=1}^N\sigma_{zi}U_F(t')|\psi(0)\rangle.
\end{equation}
But,
\begin{equation}
    \sum_{i=1}^N\sigma_{zi}\prod_{i=1}^N|-1\rangle_i=|\chi'_0\rangle
\end{equation}
where
\begin{equation}
    |\chi'_0\rangle=|1\rangle_1|-1\rangle_2\cdots|-1\rangle_N+
	|-1\rangle_1|1\rangle_2\cdots|-1\rangle_N+\cdots+
	|-1\rangle_1|-1\rangle_2\cdots|1\rangle_N
\end{equation}
giving a modified state of the environment. This gives
\begin{equation}
    |\psi^{(1)}(t)\rangle=-i\Delta U_F(t)\int_0^tdt'
	e^{i(2N-1)\frac{g^2}{\omega^2}(\omega t'-\sin(\omega t'))}
	\exp[\beta(t')a^\dagger-\beta^*(t')a]|0\rangle|\chi'_0\rangle
\end{equation}
being
\begin{equation}
    \beta(t)=\frac{g}{\omega}(e^{i\omega t}-1).
\end{equation}
The point here is that, when the environment is traced away, it does not give
any contribution as the modified environment state is orthogonal (but not
normalized) to the initial state.

Finally, we easily recognize that, in the limit $N\rightarrow\infty$, using the
Riemann's lemma, the first order correction goes to zero.

In this way, we have found a method to generate classical radiation fields of arbitrary
amplitude by unitary evolution in the thermodynamic limit.

Now, we prove that a single spin interacting with a spin-bath of two-level systems as the
one of our first example can, actually, produced decoherence. This model was
considered initially by Zurek in \cite{zur2}. We want to prove in a different
way that the limit of increasing number of spins in the bath can make the
single spin decohere directly from unitary evolution. The Hamiltonian can be
written as
\begin{equation}
    H = \lambda\tau_x\sum_{i=1}^N\sigma_{xi}
\end{equation}
being $\lambda$ an energy scale, $\tau_x$ the Pauli matrix for the single spin interacting with the
bath and $\sigma_{xi}$ the Pauli matrix for the $i$-th spin in the bath.
Now, as we are aware that the way the system is prepared is critical, we take as
an initial state as
\begin{equation}
    |\psi(0)\rangle=|\downarrow\rangle\prod_{i=1}^N|-1\rangle_i
\end{equation}
being $|\downarrow\rangle$ the initial state of the interacting spin eigenstate of $\tau_z$ 
and $|-1\rangle_i$ eigenstate of $\sigma_{xi}$
in the bath. The unitary evolution gives the exact solution
\begin{equation}
    |\psi(t)\rangle =\exp\left(iN\lambda t\tau_x\right)|\psi(0)\rangle
\end{equation}
and this yields, for the interacting spin, a coherent spin state with a parameter N.
We rewrite it, omitting the bath, as
\begin{equation}
    |\psi(t)\rangle_I=\cos(iN\lambda t)|\downarrow\rangle+i\sin(iN\lambda t)|\uparrow\rangle
\end{equation}
and we have Rabi flopping with a frequency $\Omega=2N\lambda$ that goes to
infinity in the thermodynamic limit. A meaning should be attached to the limits
\begin{eqnarray}
    \lim_{N\rightarrow\infty}\cos(Nx) \\
	\lim_{N\rightarrow\infty}\sin(Nx)
\end{eqnarray}
and this can be done if we rewrite the above as
\begin{eqnarray}
    \cos(Nx)&=&1-\int_0^{Nx}\sin(y)dy \\
	\sin(Nx)&=&\int_0^{Nx}\cos(y)dy
\end{eqnarray}
that we can reinterpret e.g. through the Abel summation to divergent integrals \cite{hardy}
\begin{eqnarray}
    \lim_{\epsilon\rightarrow 0^+}\lim_{N\rightarrow\infty}\int_0^{Nx}e^{-\epsilon y}\cos(y)dy \\
	\lim_{\epsilon\rightarrow 0^+}\lim_{N\rightarrow\infty}\int_0^{Nx}e^{-\epsilon y}\sin(y)dy,
\end{eqnarray}
where the order with which the limits are taken is important. These give $0$ and $1$
respectively that means average in time. This is, indeed, the only meaning
that can be attached  to functions having time scales of variation going to $0$.

This conclusion is fundamental for the density matrix as, in this way, turns out
to have the off-diagonal terms averaged to zero.
This is decoherence but obtained in the thermodynamic limit through unitary evolution.
In fact, one has
\begin{equation}
    \rho(t)=\exp\left(itN\lambda\tau_x\right)
	\left(
	\begin{array}{clcr}
	0 & 0 \\
	0 & 1
	\end{array}
	\right)\exp\left(-it\lambda \tau_x\right)
\end{equation}
that yields
\begin{eqnarray}
    \rho_{\uparrow\uparrow}(t)&=&\frac{1-\cos(2N\lambda t)}{2} \\
	\rho_{\uparrow\downarrow}(t) &=& i\frac{1}{2}\sin{2N\lambda t} \\
	\rho_{\downarrow\uparrow}(t) &=& -i\frac{1}{2}\sin{2N\lambda t} \\
	\rho_{\downarrow\downarrow}(t)&=&\frac{1+\cos(2N\lambda t)}{2}.
\end{eqnarray}
and applying the above argument in the limit $N\rightarrow\infty$ gives the required result
$\rho_{\uparrow\downarrow}(t)=\rho_{\downarrow\uparrow}(t)=0$ 
and $\rho_{\uparrow\uparrow}(t)=\rho_{\downarrow\downarrow}(t)=\frac{1}{2}$.

The main conclusion is that quantum mechanics can generate by itself classical
objects without interaction with an external environment. Although this cannot
be seen as decoherence, the latter can be recovered in some cases as
oscillations that are washed out in the thermodynamic limit.

The foundation of this approach is the formal analogy between the time evolution
operator in quantum mechanics and the density matrix for a thermodynamical system.
This can be obtained by a Wick rotation $t \rightarrow -i\beta$.

This, in turn, means that the thermodynamic limit can have a meaning also for
quantal evolution. The laws of motion obtained in this limit are so given, through
the Ehrenfest theorem, by the classical equations of motion without any deviation
due to quantum fluctuations. This is analogous to the laws of thermodynamics
obtained from the partition function.

This means that quantum mechanics can prove to be just the bootstrap program of the
classical world.

\end{document}